\documentclass[%
 reprint,
 amsmath,amssymb,
prb,
]{revtex4-2}

\usepackage{float}
\usepackage{comment}
\usepackage{tabularx}

\usepackage[utf8]{inputenc}
\usepackage[T2A,T1]{fontenc}
\usepackage[english]{babel}

\usepackage{color}
\usepackage{ulem}
\usepackage{xcolor}

\usepackage{xfrac}

\usepackage{gensymb}

\usepackage[caption=false]{subfig}

\usepackage{graphicx}% Include figure files
\usepackage{dcolumn}% Align table columns on decimal point
\usepackage{bm}% bold math

\begin{document}

\title{Plasmonic grating for circularly-polarized out-coupling of waveguide-enhanced spontaneous emission}

\author{Ilia M. Fradkin}
\email{Ilia.Fradkin@skoltech.ru}
\affiliation{Skolkovo Institute of Science and Technology, Bolshoy Boulevard 30, bld. 1, Moscow 121205, Russia}
\affiliation{Moscow Institute of Physics and Technology, Institutskiy pereulok 9, Moscow Region 141701, Russia}
\author{Andrey A. Demenev}
\affiliation{Institute of Solid State Physics, Russian Academy of Science, Chernogolovka 142432, Russia}
\author{Vladimir D. Kulakovskii}
\affiliation{Institute of Solid State Physics, Russian Academy of Science, Chernogolovka 142432, Russia}
\affiliation{National Research University Higher School of Economics, Myasnitskaya Street 20, Moscow 101000, Russia}
\author{Vladimir N. Antonov}
\affiliation{Skolkovo Institute of Science and Technology, Bolshoy Boulevard 30, bld. 1, Moscow 121205, Russia}
\affiliation{Physics Department, Royal Holloway, University of London, Egham, Surrey TW20 0EX, United Kingdom}
\author{Nikolay A. Gippius}
\affiliation{Skolkovo Institute of Science and Technology, Bolshoy Boulevard 30, bld. 1, Moscow 121205, Russia}

\date{\today}

\begin{abstract}
Plasmonic metasurfaces form a convenient platform for light manipulation at the nanoscale due to their specific localized surface plasmons.
Nevertheless, despite the high degree of light localization in metals, their intrinsic Joule losses are often considered prevention from applications in high-quality dielectric structures.
Here, we experimentally demonstrate that in some cases, the capabilities of plasmonic particles for light manipulation prevail over the negative impact of absorption.
We show the lattice of plasmonic nanoparticles onto a dielectric waveguide that efficiently couples the light of both circular polarizations to guided modes propagating in opposite directions. We demonstrate 80\% degree of circular polarization for the out-coupled emission of GaAs-waveguide-embedded quantum dots.
The results allow us to consider the lattice as a circular-polarization-controlled grating coupler operating at normal incidence and make this structure prospective for further implementation as an efficient coupling interface for various integrated devices.

\end{abstract}

%\keywords{Suggested keywords}%Use showkeys class option if keyword
                              %display desired
\maketitle

%\tableofcontents

\section{Introduction}

The last decade is marked with a fast progress in integrated photonics. A number of the nanophotonic elements and circuits emerged~\cite{shen2017deep,fang2015nanoplasmonic,bogaerts2020programmable}, the optical metasurfaces became the basis for many devices, and the concept of flat optics became very popular and widely applied~\cite{yu2014flat,chen2016review,kildishev2013planar}. From this prospective, there is a need of an efficient controllable coupling between the free propagating light and optical modes of the integrated structures. Many systems are designed to out-couple the radiation from integrated light sources directly~\cite{aouani2011plasmonic, wang2015broadband, ziebarth2004extracting}. But most of them are aimed at injection or extraction of the guided modes~\cite{tamir1977analysis,bates1993gaussian,mehta2017precise,taillaert2004compact,chen2010apodized,marchetti2017high,zaoui2012cost,michaels2018inverse,sacher2014wide,hong2019high,ding2014fully,taillaert2002out,taillaert2003compact,mekis2010grating} and coupling them with optical fibers~\cite{mehta2017precise,taillaert2004compact,chen2010apodized,marchetti2017high,zaoui2012cost,michaels2018inverse,sacher2014wide,hong2019high,ding2014fully,taillaert2002out,taillaert2003compact,mekis2010grating}. To achieve the high level of efficiency apodized (gradient) gratings are applied~\cite{mekis2010grating,halir2009waveguide,mehta2017precise,taillaert2004compact,chen2010apodized,marchetti2017high,sacher2014wide,hong2019high,ding2014fully} as well as some structures of advanced designs~\cite{zaoui2012cost,michaels2018inverse,sacher2014wide,hong2019high,ding2014fully}. The most useful gratings not only provide the in/out-coupling interface but offer an additional functionality such as polarization/orbital angular momentum splitting~\cite{taillaert2003compact,mekis2010grating,zhou2019ultra,zhao2019compound,nadovich2016forked}, beam focusing via concentric gratings~\cite{aouani2011plasmonic,steele2006resonant,bachman2012spiral,qi2015enhancing}, biosensing~\cite{voros2002optical,ramuz2011optical},  integration with hyperbolic metamaterials~\cite{lee2021angular,sreekanth2014large}, etc~\cite{quaranta2018recent}. Nevertheless, in many case the systems are rather complicated in operation and fabrication. As a result, the majority of the coupling systems used in practical applications still have very simple design~\cite{redding2013compact,khasminskaya2016fully,kahl2015waveguide,vetter2016cavity,lazarenko2021size,ramuz2011optical,ramuz2012transparent}.

In this work we explore the gratings of plasmonic nanoparticles. They have never been seriously considered as couplers because of the Joule heating in metals, which limits the maximum achievable coupling efficiency. Nevertheless, the localized surface plasmon resonances in relatively small plasmonic nanoparticles make them efficient antennas that rout energy rather than absorb it. Despite the plasmonic gratings are not suitable for the achievement of the recording efficiencies but they can be a very convenient platform for the design of complex metasurfaces of small dipolar particles having disproportionally strong optical responses.
These motivated a wide application of the plasmonic gratings in holography~\cite{zheng2015metasurface,ye2016spin,wei2017broadband}, SERS~\cite{kukushkin2020metastructures,fedotova2019spoof,kukushkin2017size}, biosensors~\cite{Shen2013} and many other optical metasurfaces~\cite{kravets2018plasmonic,Rajeeva2018,utyushev2021collective,vaskin2019light,kolkowski2019lattice}.

The lattice of perpendicular nanoslits in plasmonic wafer for the circular polarization-dependent excitation of surface plasmons was proposed in the pioneering paper by Capasso~\cite{lin2013polarization}. This study was followed up by a number of applications of similar structures for the so-called photonic spin-orbit coupling~\cite{bliokh2015spin,shitrit2013spin}. Most of them were also based on surface plasmons~\cite{lee2015plasmonic,zhao2019compound,mueller2014polarization,yin2018polarization,miroshnichenko2013polarization,zhao2019polarization,huang2013helicity,ding2018bifunctional,yang2020deuterogenic,chen2020polarization,xu2017polarization,lee2018ultracompact,zhang2018polarization,consales2020metasurface,Pham2018,Revah2019}, although the same principle is valid for dielectric waveguides as well~\cite{fradkin2020nanoparticle,jiang2018metasurface}. Remarkably, the dielectric structures schemes attracted much less attention in this context, although they are much more widely spear in practice.

In this paper,
we experimentally demonstrate 
effective coupling of the modes propagating in the integrated optical waveguides with circularly-polarized-light in the far-field.
Our plasmonic grating comprised of two perpendicular nanorods in a unit cell (see~Fig.~\ref{fig:1}), which couples the left- and right-hand circularly polarized light with photonic guided modes propagating in opposite directions.
We show that such grating supports circularly polarized light in the far-field for both, quantum dots emission from under it and for out-coupling of the side-coming guided modes. In the latter case, the degree of circular polarization reaches more than 80\%. Thus we have a circular-polarization-controlled grating coupler efficiently operating even for the light propagating normally to the chip. With a further development, this type of grating can be a base for the new platform for manipulating photonic guided modes in a way similar to already established techniques of surface plasmons control. The experiments pave the way for practical realization of gradient metasurfaces, splitters of circular polarizations to TE/TM modes, effective on-chip sources of circularly polarized light, and many other potential devices.

\section{Methods}

The gratings of $30\times30\mu\mathrm{m}^2$ size (see Fig.~\ref{fig:1}~(b)) were fabricated onto the 263~nm GaAs waveguide layer ($\varepsilon_{\mathrm{GaAs}}=12.42$~\cite{lobanov2015controlling}) with self-assembled InAs quantum dot (QD) layer inside (100~nm above the layer bottom). The waveguide itself lies on the AlGaAs ($\varepsilon_{\mathrm{AlGaAs}}=9.66$~\cite{lobanov2015controlling}) 1~$\mu$m-buffer atop of the GaAs. The plasmonic grating of golden nanorods is fabricated via the standard lift-off technique. The rods are made of 25~nm thick gold film with the 5~nm Al adhesion layer to GaAs.

The polarization properties of the emission of InAs QDs (see supplementary material~(B) %\ref{sup:B} 
for spontaneous emission spectrum) from a GaAs waveguide with plasmon gratings were studied at a temperature of $T=8$~K. The structure was placed in the cryostat with a cold finger. The 532~nm laser beam was focused into a spot 30~$\mu$m in diameter on the studied grating with dimensions of $30 \times 30\mu\mathrm{m}^2$ or outside it on the waveguide structure. The optical setup made it possible to measure emission spectra both in real and momentum spaces. The latter was determined from angle-resolved measurements of the QD emission spectra, $I(\hbar\omega,\Theta)$, in the cone of angles $|\theta|\lesssim10\degree$ according to the relation $k_x=2\pi/\lambda*\sin\theta$ . Therefore, all the $\hbar\omega-k_x$ maps were measured in the range $|k_x|\lesssim1.2\mu\mathrm{m}^{-1}$ ($\approx0.05*2\pi/a_x$ at $\lambda=885$~nm for the investigated gratings with a lattice period $a_x=267$~nm). Circularly polarized components of QD emission were detected with a combination of broadband polarizer and achromatic $\lambda/4$ plate.
The spatial, angular and spectral resolutions used were 2~$\mu$m, $0.50\degree$ ($\approx0.0026*2\pi/a_x$ in $k$-space), and 0.21~meV, respectively.

All the calculations conducted via Fourier modal method~\cite{tikhodeev2002} enhanced by discrete dipole approximation~\cite{fradkin2019fourier,fradkin2020nanoparticle,fradkin2020thickness}. Dipole polarizability of golden nanoparticles is calculated via finite element method (FEM)-based calculation technique~\cite{fradkin2019fourier} (see supplementary material~(A)%\ref{sup:A}
). Gold and aluminum properties are described by Johnson\&Christy~\cite{JohnsonChristy1972} and Rakic~\cite{rakic1995algorithm} optical constants correspondingly. For simplicity, GaAs wafer is not accounted in the calculations, because it does not affect the optical properties of the structure.
The details of the calculations are discussed throughout the paper and in supplementary material~(B,C).

\section{Results and Discussion}

\begin{figure}
    \centering
    \includegraphics[width=0.9\linewidth]{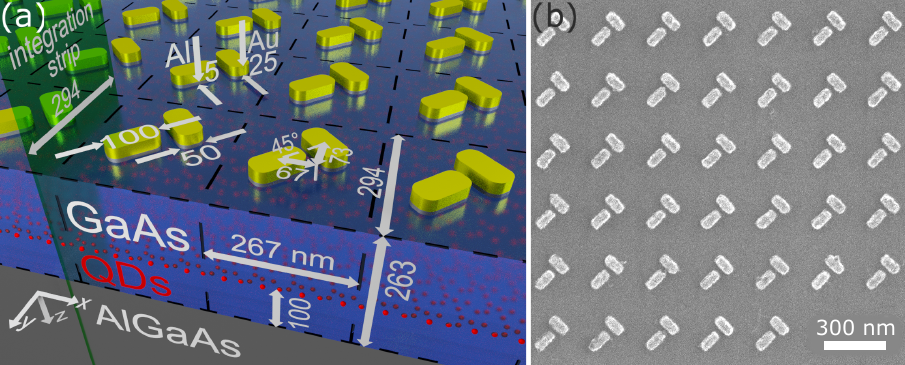}
    \caption{(a)~Schematic of the experimental structure. The GaAs 263~nm thick waveguide is grown at the top of GaAs wafer with 1~$\mu$m AlGaAs termination. Grating of the golden nanoparticles is at the top of GaAs waveguide layer. In-plane polarized QDs are embedded in the waveguide 163~nm below the surface.
    The green translucent strip indicates the area through which we calculate the guided mode flux.
    (b)~Scanning electron microscope micrograph of the plasmonic grating.
    }
    \label{fig:1}
\end{figure}

\begin{figure*}
    \centering
    \includegraphics[width=0.75\textwidth]{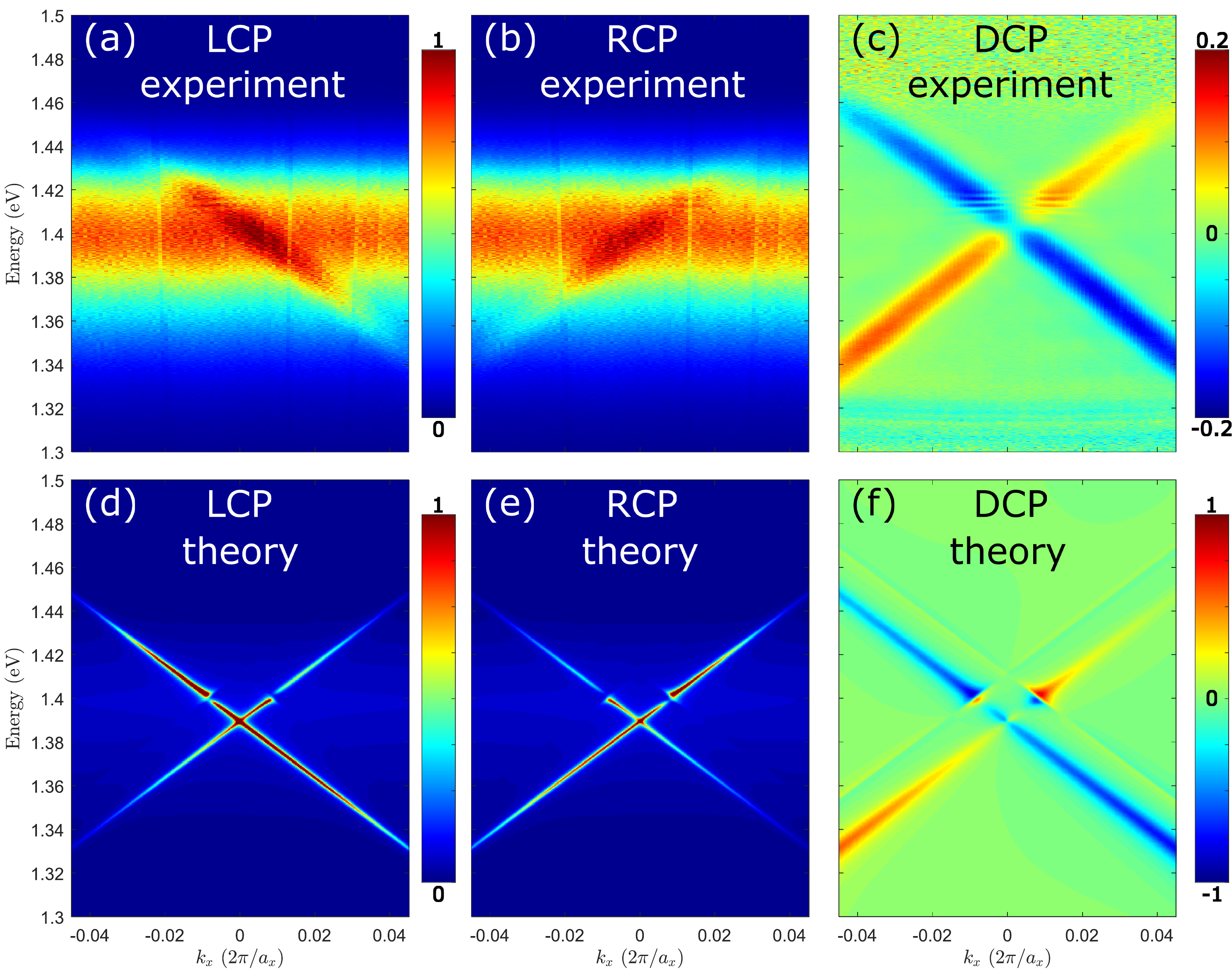}
    \caption{Experimental (a-b) and theoretical (d-e) emission spectra of QDs from under the plasmonic grating. Left-propagating TE guided mode is mostly left-hand circularly polarized (a,d), whereas right-propagating one is right-hand polarized (b,e). The degree of circular polarization (c,f) represents a cross of branches of opposite signs. The discrepancy between the experiment (a-c) and theory (d-f) is attributed both to experimental imperfections and theoretical approximations implemented in practical calculations.
    }
    \label{fig:2}
\end{figure*}

We start with probing the spontaneous emission of QDs from one of the gratings. The QDs embedded in the GaAs layer beneath the lattice are excited with the focused laser beam. Their emission spectrum modified by the grating is measured as a function of the $k_x$-projection of the wavevector ($k_y=0$). Fig.~\ref{fig:2}~(a-b) shows experimentally measured $\hbar\omega-k_x$ maps of  QDs emission for (a) left-hand, $I^\mathrm{LCP}$, and (b) right-hand, $I^\mathrm{RCP}$, circularly polarized light. The wide band of unpolarized emission ($1.38-1.42$~eV) observed for both spectra is associated with the intrinsic spectrum of QDs spontaneous emission (see supplementary material~(B)%\ref{sup:B}
). At the same time, this emission is resonantly enhanced due to the Purcell effect~\cite{purcell1995spontaneous} with the guided modes of the structure~\cite{tikhodeev2002}. Since QDs emission is mostly in-plane polarized, it couples efficiently only to transverse electric, TE
modes. Therefore, we observe 
only TE modes in experimental spectra~in~Fig.~\ref{fig:2}~(a,b).
What is most important, plasmonic grating scatters out the left-propagating mode (negative incline) with a left-hand circular polarization,~Fig.~\ref{fig:2}~(a), and the right-propagating one (positive incline) with a right one,~Fig.~\ref{fig:2}~(b). As a result, the degree of circular polarization of the emitted light $\mathrm{DCP}=\frac{I^\mathrm{RCP}-I^\mathrm{LCP}}{I^\mathrm{RCP}+I^\mathrm{LCP}}$ 
comprises two bright lines of TE modes that have opposite circular polarization in the far-field, red and blue lines in Fig.~\ref{fig:2}~(c). Although the guided modes are clearly distinguished by their polarization, the value of DCP ($\approx16\%$ in a peak) is rather small due to the relatively large magnitude of the background, non-resonant emission from QD.
Interestingly, it might seem that the maps of emission in complementary polarizations, panels (a-b), are the mirror reflections of each other, but the symmetry of the structure does not provide this effect since the $x=\mathrm{const}$ plane is neither the reflection nor the glide reflection plane of the considered crystal, see Fig.~\ref{fig:1}. Also, it is worth mentioning that the pale hardly distinguishable horizontal lines in Fig.~\ref{fig:2}~(c) at $\approx1.31$~eV correspond to the $y$-axis propagating guided modes. These modes are intentionally lowered in energy by the relatively large $y$-axis period $a_y=294$~nm in order to prevent their hybridization with the studied resonances.

Accurate calculations of the emission spectra for a 2D lattice of plasmonic nanoparticles is a very challenging problem. However, dipole approximation~\cite{fradkin2020nanoparticle} makes it possible to carry out qualitative estimations illustrated in Fig.~\ref{fig:2}~(d-f). In the framework of this approach, we describe each nanorod by a dipole polarizability tensor (see supplementary material~(A)%\ref{sup:A}
), which is actually not comprehensive in our case. Indeed, the nanorod's length makes almost a half of the guided mode wavelength, with the period $a_x=267$~nm in $x$-direction, and even larger fraction of the gold/GaAs-interface surface plasmon's wavelength. Also, the sharp edges of the particle might lead to the extremely localized edge plasmons that are extremely sensitive to the shape of the edge~\cite{gramotnev2010plasmonics} and the optical properties of the materials on the scale of several nanometers~\cite{ford1984electromagnetic}. Therefore, the nanorods optical response is too complex to be accurately described by the dipole approximation since they have a complex multipolar responses to the electric field and its gradients. Nevertheless, the dipole contribution is still dominating,
and it can be used for qualitative analysis.

Fig.~\ref{fig:2}~(d-f) show 
the emission spectra and corresponding DCP calculated in dipole approximation. The emission intensity is given in arbitrary units 
so they can be compared with experimental spectra (Fig.~\ref{fig:2}~(a-b)) only qualitatively. Nevertheless, we are still able to judge on the similarities and differences in corresponding spectra. Generally, the calculations match the experimental results, but there are several significant discrepancies as well. 
Theoretical estimations for the infinite grating demonstrate much more pronounced and narrow resonances. Moreover, in addition to the left (right)-propagating TE modes with left (right)-hand circular polarization observed in the experiment, there is also a weak emission of the oppositely-propagating modes, which results in $\sim 50-70\%$ degree of circular polarization, see Fig. 2 (d-f).

Discrepancies between the theory and experiment are mainly related to the finite size of the fabricated structures. In particular, the diffraction on $D=30\mu$m rectangular aperture gives the estimation for the mode linewidth in momentum space $\delta k_x \approx 2\pi/\lambda\cdot \lambda/D\approx 0.09 \cdot2\pi/a_x$. This value is much larger than theoretical predictions for infinite lattice and therefore determines the true width of the resonances. The weak experimental, spontaneous emission rate enhancement is also associated with the finite size of the grating, non-tabular permittivity of golden nanobars and Al adhesion layer, edge, and other effects. Another obvious peculiarity is the anticrossing behavior of TE and TM modes ($\approx 1.4$~eV), which exists in all computational spectra but is absent in the experimental ones. This is explained by the small value of the coupling constant compared to the large linewidth determined by the finite lattice size. In this way, the compliance between theory and experiment can be considered satisfactory.

In fact the core optical properties of the structure might be explained qualitatively even without any calculations. Although plasmonic nanoparticles strongly scatter light, they are weakly coupled with photonic guided modes of the dielectric structure. This means that outside the narrow energy range of modes anticrossing, plasmonic nanoparticles just scatter out the guided modes and almost do not change the dispersion and fields distribution. Then the QD emission might be qualitatively explained by the two-step process. Firstly, the QDs resonantly excite the guided modes of the dielectric structure due to the Purcell effect~\cite{purcell1995spontaneous}. Then, these modes are scattered out by plasmonic grating. We assume the nanobars to be the uniaxial scatterers (extreme anisotropy) and there are two sublattices, which scatter out light independently on each other. Each of the sublattices couples linearly-polarized TE modes, propagating in both directions along the $x$-axis, to the far-field plane-waves of perpendicular polarizations. A $\pm\pi/2$ 
phase lag arises due to the quarter-period $x$-axis shift (for the centers of the particles), which produces circularly-polarized radiation for the oppositely propagating guided modes~\cite{lin2013polarization}. The fact that none of the axes of the particles dominates in optical response results in weakening the effect (see supplementary material~(A)).

\begin{figure}
    \centering
    \includegraphics[width=0.8\linewidth]{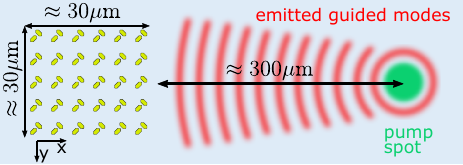}
    \caption{Schematic of the plasmonic lattice operating as a grating out-coupler. QDs excited at a distance of $300\mu$m from the grating resonantly re-emit light in the guided modes, which reach the grating coupler in the form of an almost plane wave. The grating subsequently scatter out radiation to the far-field.
    }
    \label{fig:3}
\end{figure}

\begin{figure*}
    \centering
    \includegraphics[width=0.75\textwidth]{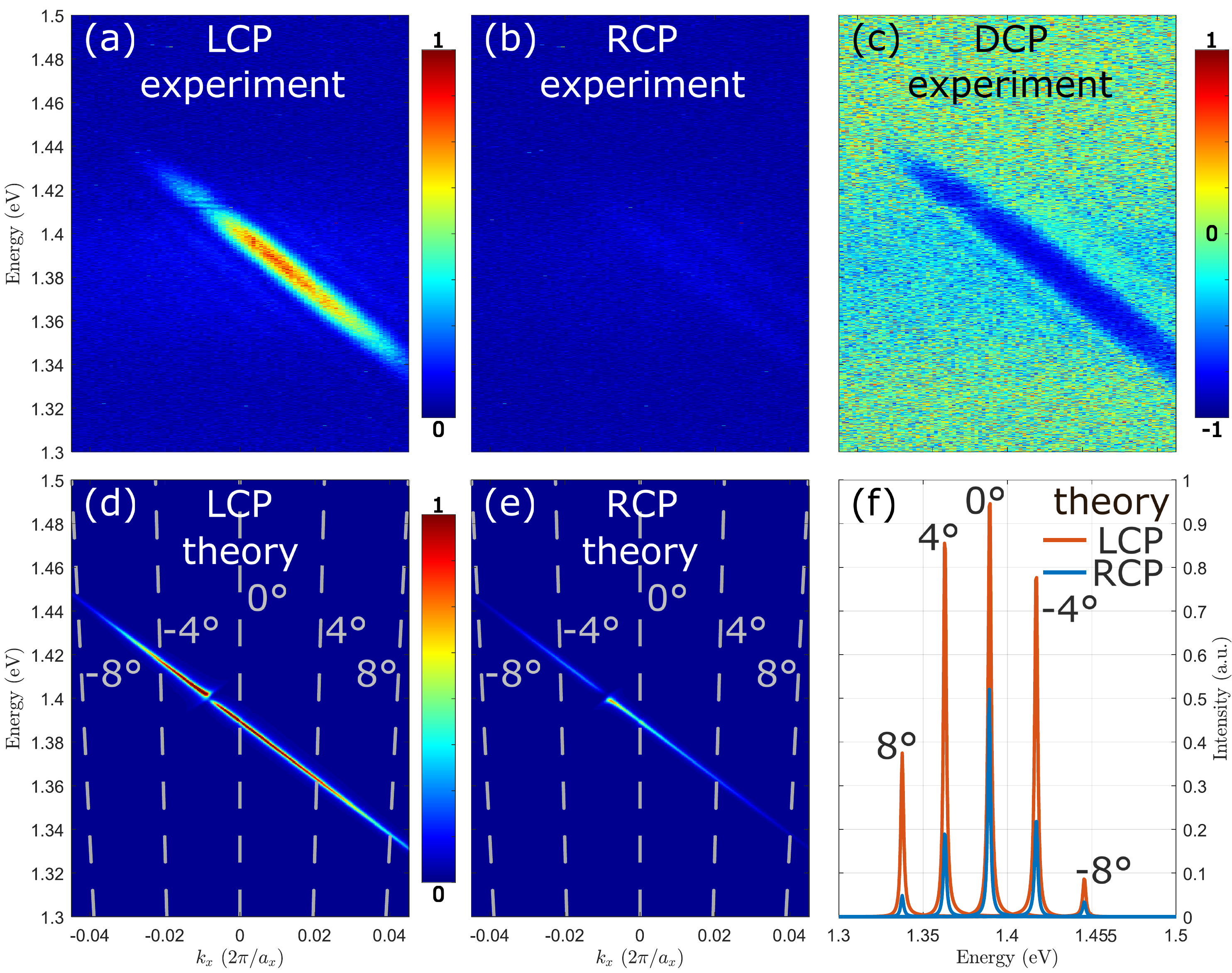}
    \caption{Experimental spectra (a-b) and their theoretical estimations (d-e) for the light out-coupled by the grating coupler in the scheme from Fig.~\ref{fig:3}. Most of the radiation is left-hand circularly polarized (a,d), which results in approximately 80\% degree of circular polarization (c). Theoretical estimations of emission spectra for different propagation angles also demonstrate the domination of one of the circular polarizations over another, which is clearly seen from the energy spectra for several fixed angles (f). 
    }
    \label{fig:4}
\end{figure*}

As it was already discussed, the maximum DCP is mostly limited by the high level of the background emission, which is comparable with the resonant emission. It is possible to increase DCP by suppressing all the dissipation channels to make the resonances more pronounced. Alternatively one can filter out the background emission, which technically is much simpler.
In order to do that, we excite QDs inside the uncovered waveguide area at a distance of $\approx 300~\mu$m from the out-coupling grating, see Fig.~\ref{fig:3}. Emitting QDs resonantly excite TE modes propagating in all directions.
The guided mode reaches the grating with almost a plane front. In this way, the dielectric waveguide naturally filters out the background emission.
The corresponding out-coupling spectra
contain only left-propagating TE mode, see Fig.~\ref{fig:4}~(a-b). It is important that the TE mode is mostly pronounced in left-hand circular polarization of far-field, in accordance with the previous measurements.
The emission maps prove that the background field is strongly suppressed, so that the DCP is raised up to $\approx80\%$, see Fig.~\ref{fig:4}~(c).  
We clearly observe the fine effect of multiple replicas of the original mode, which is 
associated with diffraction on 
the patch of the grating.

The out-coupling of the guided modes by a finite-size grating is a challenging task for simulation. Indeed, the grating is too large ($30\times30\mu\mathrm{m}^2$) to treat the problem in a manner of scattering by ordinary particles. At the same time, the grating cannot be considered an infinite one since the out-coupling is a fundamentally edge-assisted process. For this reason, we developed a specialized qualitative theoretical approach to verify experimental results.

In order to ease the estimations, we apply the Lorentz reciprocity principle. Instead of the original problem of left-propagating guided wave out-coupling, we consider excitation of the right-propagating TE mode by the light of both circular polarizations. In this way, the flux of the guided mode coming out from the illuminated grating-covered area is proportional to the intensity of the correspondingly polarized light out-coupled by this lattice under the guided mode incidence.
Although the in-coupling efficiency is just as challenging to calculate as the out-coupling one, it might be easily estimated with accuracy of constant factor.
Indeed, the plasmonic lattice is weakly-coupled to the photonic guided mode, and therefore the reflection of the guided wave from the boundaries of the finite grating is also small. Moreover, guided waves propagating in opposite directions are uncoupled with each other (except for the narrow energy band of anticrossing). Altogether this means that the oppositely-propagating guided modes almost do not interact with each other, and their excitation, scattering, absorption, and other kinetics might be considered independently.
In particular, the flux of the guided modes is balanced mainly by the in-coupling of the incident light, backward out-coupling, and absorption in the metal. For the infinite lattice, we deal with a stationary process when the guided mode energy flux does not vary from one period to another. It is important, that within our assumptions, the flux of right-propagating mode over some $x=\mathrm{const}$ section (see the green strip in Fig.~\ref{fig:1}) is entirely harvested by the lattice from the left half-space. Moreover, most of the flux is attributed to the in-coupling near the edge of the semi-infinite lattice since the guided mode excited at a far distance from the edge completely extincts, reaching the section of consideration. Therefore, guided modes excitation by such large gratings as we study experimentally ($30\mu\mathrm{m}\times30\mu\mathrm{m}$) might be associated with excitation by semi-infinite lattice and, in turn, with modes propagating in the bulk of the infinite lattice. The last quantity can easily be computed due to the periodicity of the structure.
One can virtually illuminate the structure with the light of different polarizations and compute the flux of the generated modes in the $x=\mathrm{const}$ section.
The details of the calculation procedure, which is conducted in dipole approximation for the lattice and resonant approximation for the modes, are discussed in supplementary material~(B,C).

Our theoretical estimations can be compared with experiments only qualitatively. Nevertheless, as we can see from Fig.~\ref{fig:4}~(d-e), the results are indeed very similar to the experimental ones. In the resonance light of left-hand circular polarization is approximately one order of magnitude more intense than of right-hand polarization. We even observe the deep for the TE and TM modes intersection point, which is notable on the experimental measurements.

Unfortunately, our computational technique does not provide us an opportunity to calculate the degree of circular polarization for arbitrary energy and angle. The reason is that in most cases, we are out of the resonant approximation, and the resulting degree of polarization is totally determined by the non-resonant effects that are beyond the scope of the model. Moreover, the intensity for most of the points is very low, and therefore even minor corrections can totally change the effect. For this reason, we only consider the energy spectrum of the out-coupling intensity of the light of both circular polarizations for several angles of propagation (see Fig.~\ref{fig:4}~(f)). The graph shows the lineshape of the light intensity in the corresponding sections of maps from the panels (d)~and~(e). We see that the height of the peaks is mostly determined by the energy of the guided mode and the corresponding emissivity of the QDs. At the same time ratio of the peaks height also varies depending on the energy of the modes and the proximity of their hybridization points.

The effects observed in our study show that the prominent effect of circular-polarization controlled coupling can be effectively implemented not only for the manipulation of surface plasmons but also photonic guided modes. We already can excite the unidirectional guided wave even by the normally incident light and control the direction by the sign of polarization. Vice versa, it is also possible to out-couple either the emission or the guided modes in a certain circular polarization. The presented results pave the way to further design of the plasmonic couplers with optimized characteristics. As prospective directions, we consider optimizing the collecting efficiency and development of gradient couplers adapted for beams of a finite section, simultaneous coupling of TE and TM modes. In practice, it is vital to adapt the gratings for the purposes of specific devices, such as to surround some integrated source of light in the best way to harness most of its emission. Another promising direction is the realization of multifunctional structures that combine the coupling itself with some auxiliary purposes such as routing guided modes, splitting the waves by polarization, simultaneous excitation of several guided modes in different directions, or sensing.

\section*{Conclusion}
In this paper, we experimentally study the ability of plasmonic lattices to out-couple the photonic guided modes into circularly polarized light on an example of GaAs waveguide with InAs QDs. We measured QDs emission from under the lattice and showed two resonant branches of the oppositely propagating modes with complementary circular polarization in the far-field. This effect allowed us to demonstrate the out-coupling of the TE mode generated by the spatially detached source and reach the $80\%$ degree of circular polarization. Theoretical estimations based on the dipole approximation for the plasmonic nanoparticles explain the obtained results and prove their applicability for the design of prospective structures. Our results show that plasmonic lattices appear to be convenient-in-design and effective-in-practice interfaces between the photonic guided modes and the far-field. In this way, plasmonic lattices might be implemented as multifunctional gratings, elements of on-chip light sources, routers, and many other integrated devices.

\section{Supplementary Material}

See supplementary material for the polarizability of plasmonic nanoparticles, details of the emission spectra calculations, and computation of guided modes amplitudes.

\section*{Acknowledgements}
The work of I.M.F. was supported by the Foundation for
the Advancement of Theoretical Physics and Mathematics “BASIS.”
The work of N.A.G. was supported by the Russian Foundation for Basic Research (Grant No. 18-29-20032).
The work of V.D.K. and A.A.D. was supported by the Russian Science Foundation (project 19-72-30003) .
The work of V.N.A. was supported by Engineering and Physical Sciences Research Council (EPSRC) Grant No. EP/T004088/1.
The authors acknowledge S.A. Dyakov and S.G. Tikhodeev for fruitful discussions and V.V.Ryazanov for valuable contribution.

I.M.F. designed the structures, conducted theoretical computations, prepared the manuscript. A.A.D. measured the optical properties of the structures. V.N.A. fabricated the samples. V.D.K. and N.A.G supervised the study. All authors contributed to the discussions and commented, reviewed, and edited on the paper.

\newpage

\begin{widetext}
\appendix

\section{Polarizability of plasmonic nanoparticles}
\label{sup:A}
As it was already discussed in the paper, golden nanoparticles are deposited onto a 5-nm adhesion layer of aluminum. Au thickness is 25~nm, length and width of nanobar are 100~nm and 50~nm correspondingly. The rounding radius that appears in fabrication is approximately 20~nm. Plasmonic gratings lie onto the GaAs layer of very high permittivity $\varepsilon_{\mathrm{GaAs}}=12.42$. This leads to a significant concentration of electromagnetic fields in a few nanometers around the particles' edges. Consequently, as we discussed before, the dipole approximation is not enough to accurately describe their optical properties. What is also important, in this configuration, optical properties of the whole particle are mostly determined not by the golden part but by the thin 5-nm adhesive layer. Moreover, these properties are very sensitive to the shape of the particle's edge, material's response at the scale of faces, oxidation processes on the sides, and so on. Experimental parameters of all these peculiarities are unknown, which is another reason for the impossibility of modeling the structure accurately.

In this way, we just formally calculate dipole polarizability of particles \cite{fradkin2019fourier} with square corners (90$\degree$), without any oxidation layers, and describe gold~\cite{JohnsonChristy1972} and aluminum~\cite{rakic1995algorithm} by conventional optical constants. The resulting components of the polarizability tensor are depicted in Fig.~\ref{fig:S1}.

\begin{figure}[h!]
    \centering
    \includegraphics[width=0.7\textwidth]{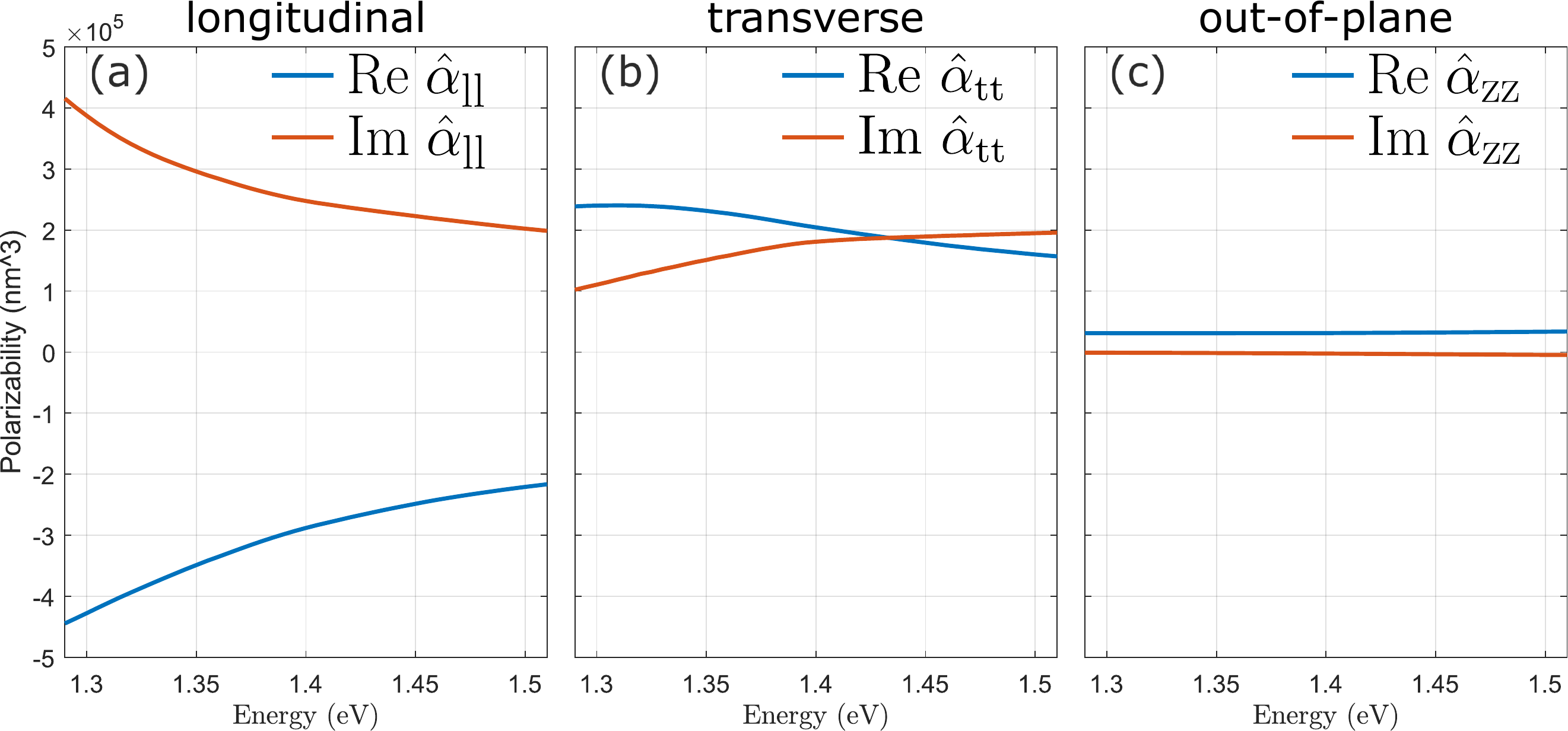}
    \caption{Spectra of longitudinal (a), transverse (b), and out-of-plane (c) components of plasmonic nanoparticle polarizability. The particle in the calculations includes not only its golden part but also the Al adhesion layer.}
    \label{fig:S1}
\end{figure}

\section{Details of practical calculations}
\label{sup:B}

As it was discussed in the study, the direct calculation of the quantities considered in the figures is rather complicated. Therefore, we employ techniques based on the Lorentz reciprocity principle.

We start with emission spectrum calculation for Fig.~\ref{fig:2}. First of all, we should keep in mind that all the in-plane polarized quantum dots distributed in the horizontal section of the waveguide emit light incoherently, and therefore, we should sum up their intensities in the far-field. Let us try to calculate the intensity of the right- or left-hand circularly polarized light radiated out from the structure with corresponding wavevector $I^{\mathrm{RCP/LCP}}_{\mathrm{far-field}}(k_x,\mathbf{r}_{\mathrm{QD}})$ by some unitary point emitter having averaged in-plane polarization. Not going into details, according to the Lorentz reciprocity principle this quantity is proportional to the intensity of the in-plane components of the electric field generated by the wave of corresponding circular polarization incident with the opposite wave vector value $I^{\mathrm{RCP/LCP}}_{\mathrm{far-field}}(k_x,\mathbf{r}_{\mathrm{QD}})\propto \left|E_{\mathbf{r}_{\mathrm{QD}}}\right|^2_{xy}(-k_x,\mathrm{RCP/LCP},\mathrm{far-field})$. In this way, one single calculation of the field distribution in the structure illuminated by the certain circular polarization allows us to average the field intensities at the position of all the incoherent sources and find their overall emission in the corresponding direction and polarization. 

Finally, we should mention that illumination of the structure for different energies and angles by unitary-intensity light provides us with a spectrum of unitary emitters. Nevertheless, in order to reproduce experimental data, we need to multiply the resulting spectra by the spontaneous emission spectrum of quantum dots in bulk GaAs $P^{\mathrm{GaAs}}(E)$. This quantity can be estimated either estimated quantum mechanically or measured experimentally. In practice, we measure the emission of quantum dots from inside the uncovered part of the waveguide $P^{\mathrm{GaAs}}_{\mathrm{waveguide}}(E,k_x)$. This quantity is not strictly the same, which we really need, but as you can see from Fig.~\ref{fig:S2}~(a), there are no any energy or angle-dependent features related to the waveguide itself. Therefore, we can assume that the bulk emission of quantum dots is proportional to the experimentally measured spectra from the waveguide with high accuracy. In order to reduce the experimental noise, we first average the measured spectrum over the propagation angle and finally smooth the resulting spectrum with a Gaussian filter of $\sigma_E=0.005$~eV (see Fig.~\ref{fig:S2}~(b)).

\begin{figure}[h!]
    \centering
    \includegraphics[width=0.8\textwidth]{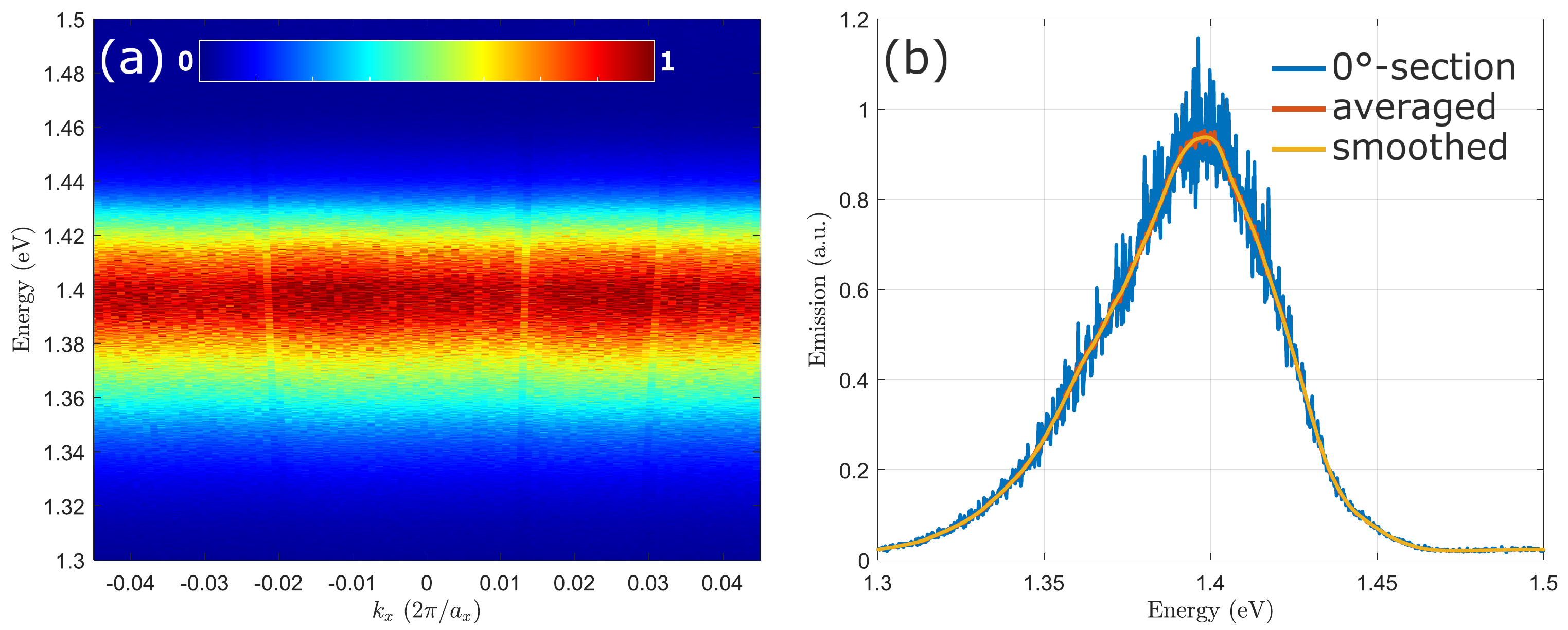}
    \caption{ (a) Spontaneous emission spectrum for the quantum dots from the uncovered area of the waveguide ($k_x$~is measured in~$2\pi/a_x$ units only for the commensurability with other graphs). (b) The energy spectrum of the quantum dots for the normal angle, spectrum averaged over the measured angles and one smoothed with a Gaussian filter.}
    \label{fig:S2}
\end{figure}

Calculations for the structure from the Fig.~\ref{fig:3} that are presented in Fig.~\ref{fig:4} were generally described in the main paper. The original idea is to apply the Lorentz reciprocity to the source that is located somewhere in the waveguide very far from the grating. In this way, this source is effectively coupled to the far-field (in terms of the interaction with a second source from the Lorentz principle) only through the TE guided mode, which it effectively excites. Therefore, the intensity of the out-coupled light $I^{\mathrm{RCP/LCP}}_{\mathrm{far-field}}(k_x,\Leftarrow\mathrm{TE\ mode})$ is associated with the flux of the right-propagating TE mode excited by the incident wave of the same circular polarization $I^{\mathrm{RCP/LCP}}_{\mathrm{far-field}}(k_x,\Leftarrow\mathrm{TE\ mode})\propto I^{\mathrm{TE\ mode}}_{\Rightarrow}(-k_x,\mathrm{RCP/LCP},\mathrm{far-field})$. Nevertheless, in order to describe the experimentally measured quantity, we need to multiply the corresponding flux by the spectrum of the quantum dots emission into TE guided mode $P^{\mathrm{GaAs}}_{\mathrm{TE\ mode}}(E)$. In principle, it is possible to calculate this quantity, knowing the emission rate in the bulk $P^{\mathrm{GaAs}}(E)$, but in practice, their shapes are almost the same, and they can be considered as equivalent ones when we speak in terms of arbitrary units.

As it was discussed in the paper, in practice, we calculate the flux of the guided mode through some section of the infinite lattice (see the green strip in Fig.~\ref{fig:1}~(a)). In this way, if we generate the incident wave having an energy flux density of $1$~W per unit cell $a_xa_y$ then we can easily calculate the guided mode flux through the strip of $a_y$ width in Watts and compare them. The technical details of the calculations are discussed in the next section. Here, we observe the results and see that the mode energy flux is rather high and reaches the value of 20 for the $8\degree$ angle of left-hand polarized light incidence. This is an interesting characteristic of the structure, which demonstrates the high quality of its resonances. Multiplication of these graphs by the envelope of the quantum dots emission spectrum (see Fig.~\ref{fig:S1}~(b)) gives us the graph from the paper (see Fig.~\ref{fig:4}~(f)).

Nevertheless, we should remember that all these calculations are conducted within dipole approximation, which is not ideal for the considered structure.

\begin{figure}[h!]
    \centering
    \includegraphics[width=0.45\textwidth]{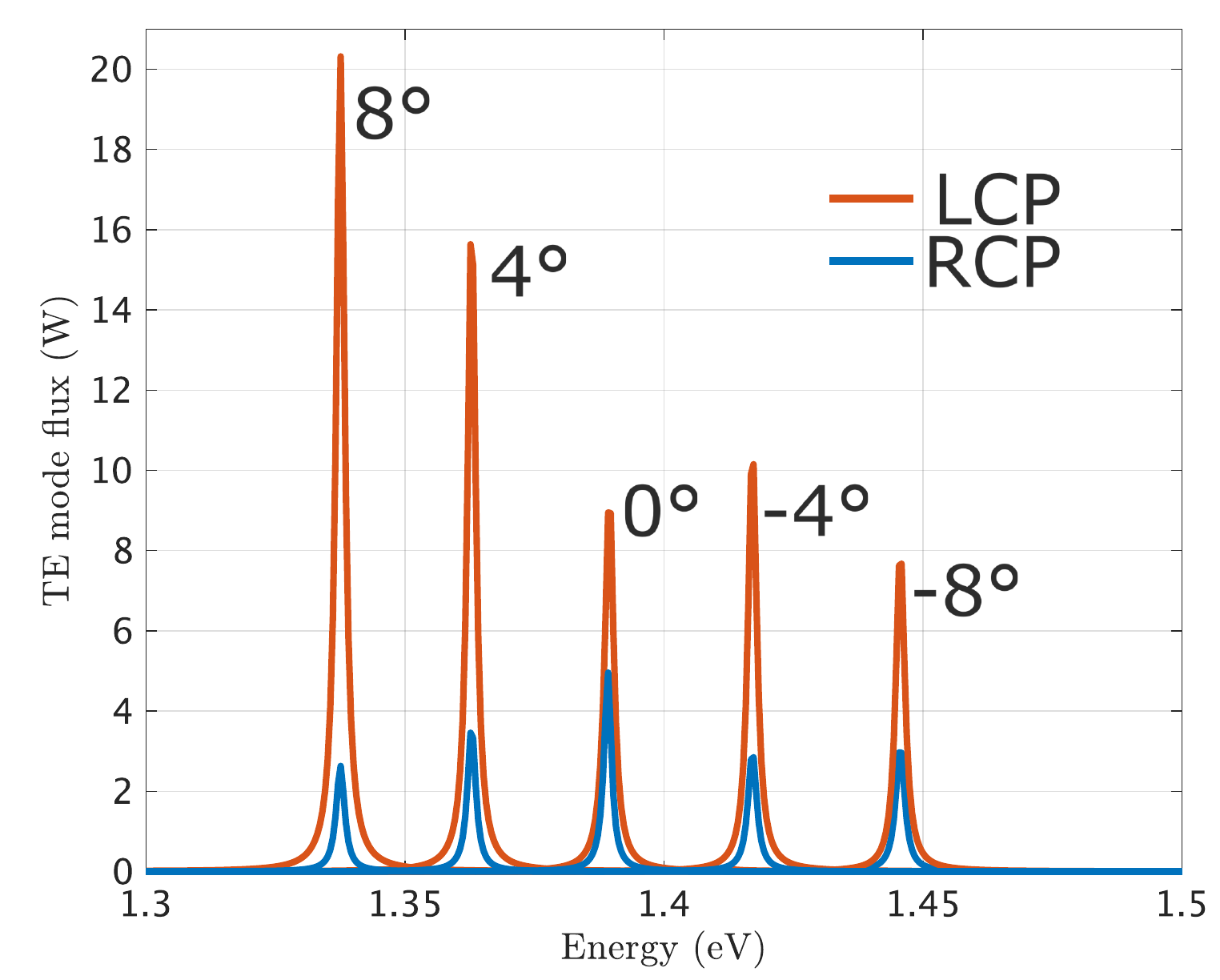}
    \caption{Energy flux of the right-propagating TE mode per each $y$-axis period, $a_y$. The structure is illuminated by a 1~W per unit cell light beam of different circular polarization at different angles of incidence.}
    \label{fig:S3}
\end{figure}

\section{Guided mode amplitude calculation}
\label{sup:C}

The main function of the plasmonic grating coupler is either to excite the guided modes under the external illumination or vice versa scatter out the guided mode in the far-field. Therefore, it is extremely convenient to have the opportunity to calculate the amplitude of the guided mode excitation and the energy flux which it generates.
Several approaches can be implemented for resonant modes consideration. For example they can be considered as quasinormal modes by different approaches~\cite{sauvan2013,lalanne2018,akimov2011optical,weiss2017analytical,gippius2010resonant}
Here, we apply slightly another approach, which is especially convenient for metasurfaces comprised of weakly-coupled slab waveguide and nanoparticles grating. In particular, the structure that we consider consists of a waveguide and a lattice of point electric dipoles. Therefore, if we consider $x=\mathrm{const}$ section of the structure that does not contain any of the dipoles (see the green strip in Fig.~\ref{fig:1}) then we deal with an ordinary waveguide in this section. It is well known, that in this case, the field in the section can be easily expanded throw the guided modes (strictly speaking solutions with $e^{i k_x x}$ $x$- dependence) propagating in both directions~\cite{vainshtein1988} in the following way:
\begin{equation}
    \begin{pmatrix}
         \mathbf{E}^{\mathrm{tot}}(x,y,z)\\
          \mathbf{H}^{\mathrm{tot}}(x,y,z)
    \end{pmatrix}=\sum_s C_s     \begin{pmatrix}
         \mathbf{E}_s(z)\\
          \mathbf{H}_s(z)
    \end{pmatrix}e^{i k^s_x x}.
\end{equation}
As long as we consider only waves propagating in the $x$ direction, the field of corresponding modes does not depend on $y$ - $\mathbf{E}_s = \mathbf{E}_s(z)$, $\mathbf{H}_s = \mathbf{H}_s(z)$.

Corresponding amplitudes $C_s$ of the $s$-th mode can be found as follows~\cite{vainshtein1988}:
\begin{equation}
    C_s =\frac{\int[\mathbf{E}^{\mathrm{tot}}\mathbf{H}_{-s}]-[\mathbf{E}_{-s}\mathbf{H}^{\mathrm{tot}}]d\mathbf{S}}{N_s}= \frac{\int[\mathbf{E}^{\mathrm{tot}}\mathbf{H}_{-s}]-[\mathbf{E}_{-s}\mathbf{H}^{\mathrm{tot}}]d\mathbf{S}}{\int[\mathbf{E}_s\mathbf{H}_{-s}]-[\mathbf{E}_{-s}\mathbf{H}_s]d\mathbf{S} },
\end{equation}
where square brackets, $[]$, denote the vector product, integration is conducted over the waveguide section, and $N_s$ is the norm of the eigenmode. In principle, this formula is sufficient for practical calculations. Nevertheless, not all the computational approaches, such as the Fourier modal method, are suitable for integrating over the $z$-axis. However, we can strongly simplify it by computing the integral analytically for separate Fourier harmonics of the field.

Since we operate in Fourier space, it is convenient to represent electromagnetic fields $\mathbf{E}^{\mathrm{tot}}(x,y,z)$ and $\mathbf{H}^{\mathrm{tot}}(x,y,z)$ that we would like to expand over the resonant modes through the Fourier harmonics:

\begin{equation}
    \mathbf{E}^{\mathrm{tot}}(x,y,z) = \sum_j \mathbf{E}^{\mathrm{tot}}_j(z) e^{i(k_x+\mathbf{g}_j\mathbf{r})},\quad     \mathbf{H}^{\mathrm{tot}}(x,y,z) = \sum_j \mathbf{H}^{\mathrm{tot}}_j(z) e^{i(k_x+\mathbf{g}_j\mathbf{r})},
\end{equation}
where the $y$-component of the incident wave $\mathbf{k}$-vector is equal to zero $k_y=0$. In this representation, we obtain the expression for $C_s$:
\begin{multline}
    C_s = \frac{\int\left[\left(\sum_j\mathbf{E}^{\mathrm{tot}}_j(z) e^{i(k_xx+\mathbf{g}_j\mathbf{r})}\right)\mathbf{H}_{-s}(z)e^{ik^{-s}_xx}\right]-\left[\mathbf{E}_{-s}(z)e^{ik^{-s}_xx}\left(\sum_j\mathbf{H}^{\mathrm{tot}}_j(z) e^{i(k_xx+\mathbf{g}_j\mathbf{r})}\right)\right]d\mathbf{S}}{N_s}=\\
    \frac{\sum_{j|g_{y,j}=0} e^{i(k_xx+\mathbf{g}_{j}\mathbf{r}+k^{-s}_xx)}a_y\left(\int\left[\mathbf{E}^{\mathrm{tot}}_j (z) \mathbf{H}_{-s}(z)\right]-\left[\mathbf{E}_{-s}(z)\mathbf{H}^{\mathrm{tot}}_j(z) \right]dz\right)}{N_s},\label{Sup:amplitude}
\end{multline}
where only the harmonics having $g_{y,j}=0$ make the non-zero contribution.
Each integral in the remaining sum can be easily calculated analytically since the $z$-dependence of both guided mode and Fourier harmonics of the total field are trivial. Outside the waveguide, they are both represented by evanescent waves, whereas inside it, the behavior is sine-cosine-like or hyperbolic sine-cosine-like. In this way, we need to compute the fields only for several $z=\mathrm{const}$ sections to recover and integrate the fields everywhere analytically.

In order to obtain the $z$-dependence of the electromagnetic fields in a certain layer, we need to find the amplitudes of up-and down-wards propagating plane waves from the in-plane components of electric $\mathbf{E}$ and magnetic $\mathbf{H}$ fields in some $z$-section. First, we need to recall the polarization of the plane waves over which we expand the total field:

\begin{equation}
    \mathbf{E}_X^\pm=\begin{pmatrix}1\\0\\\mp\frac{k_x}{k_z}
    \end{pmatrix},\quad \mathbf{H}_X^\pm=\left.\frac{1}{k_0}\begin{pmatrix}\mp\frac{k_xk_y}{k_z}\\\pm\frac{k_x^2}{k_z}\pm k_z\\-k_y
    \end{pmatrix}\right|_{k_y=0}=\frac{1}{k_0}\begin{pmatrix}0\\\pm\frac{k^2}{k_z}\\0
    \end{pmatrix},
\end{equation}

\begin{equation}
    \mathbf{E}_Y^\pm=\left.\begin{pmatrix}0\\1\\\mp\frac{k_y}{k_z}
    \end{pmatrix}\right|_{k_y=0}=\begin{pmatrix}0\\1\\0
    \end{pmatrix},\quad \mathbf{H}_Y^\pm=\left.\frac{1}{k_0}\begin{pmatrix}\mp\frac{k_y^2}{k_z}\mp k_z\\\pm\frac{k_xk_y}{k_z}\\k_x
    \end{pmatrix}\right|_{k_y=0}=\frac{1}{k_0}\begin{pmatrix}\mp k_z\\0\\k_x
    \end{pmatrix},
\end{equation}
where $\pm$ signs indicate the direction of propagation of the wave and $X/Y$ subscript indicates the polarization of the electric (not magnetic!) field of the corresponding wave. In the upwards/downwards going waves representation electromagnetic field can be found as follows:

\begin{equation}
    \mathbf{E}(z)e^{ik_xx}=\sum_{\pm,X/Y}A_{X/Y}^\pm\mathbf{E}_{X/Y}^\pm e^{i(k_xx\pm k_z(z-z_0))},\quad    \mathbf{H}(z)e^{ik_xx}=\sum_{\pm,X/Y}A_{X/Y}^\pm\mathbf{H}_{X/Y}^\pm e^{i(k_xx\pm k_z(z-z_0))},
\end{equation}
where $\mathbf{E}$ and $\mathbf{H}$ correspond either to the guided mode or to the certain harmonic of the total field.
In this way, solving the simple system of equations, we easily determine the amplitudes from the in-plane components of electromagnetic fields:

\begin{equation}\left\{
    \begin{array}{l}
         E_x(z_0) = A_X^++A_X^-  \\
         H_y(z_0) = \left(A_X^+-A_X^-\right)\frac{k^2}{k_zk_0} 
    \end{array}\right.;
\end{equation}

\begin{equation}
    \left\{
    \begin{array}{l}
        A_X^+=\frac{1}{2}\left(E_x(z_0)+\frac{k_zk_0}{k^2}H_y(z_0)\right)\\ A_X^-=\frac{1}{2}\left(E_x(z_0)-\frac{k_zk_0}{k^2}H_y(z_0)\right)
    \end{array}\right..
\end{equation}
In a similar way, we can separately obtain the expressions for $y$-polarization:

\begin{equation}\left\{
    \begin{array}{l}
         E_y(z_0) = A_Y^++A_Y^- \\
         H_x(z_0) =\left(-A_y^++A_y^-\right)\frac{k_z}{k_0}
    \end{array}\right.;
\end{equation}

\begin{equation}
    \left\{
    \begin{array}{l}
        A_Y^+=\frac{1}{2}\left(E_y(z_0)-\frac{k_0}{k_z}H_x(z_0)\right)\\ A_Y^-=\frac{1}{2}\left(E_y(z_0)+\frac{k_0}{k_z}H_x(z_0)\right)
    \end{array}\right..
\end{equation}

Now, we calculate the integral for arbitrary different plane waves employing the derived expansion.

\begin{multline}
    \int[\mathbf{E}_1\mathbf{H}_2]\hat{x}dz = \int
         (E_yH_z-E_zH_y)e^{i(k_{x1}+k_{x2})x}dz=\\
         \int
         [(A_{y1}^+e^{ik_{z1}z}+A_{y1}^-e^{-ik_{z1}z})\frac{k_{x2}}{k_0}(A_{y2}^+e^{ik_{z2}z}+A_{y2}^-e^{-ik_{z2}z})-\\-\frac{k_{x1}}{k_{z1}}(-A_{x1}^+e^{ik_{z1}z}+A_{x1}^-e^{-ik_{z1}z})\frac{k^2}{k_{z2}k_0}(A_{x2}^+e^{ik_{z2}z}-A_{x2}^-e^{-ik_{z2}z})]e^{i(k_{x1}+k_{x2})x}dz
\end{multline}

This shows that $X$ and $Y$ polarized waves make independent contributions, and we can consider them separately. We consider only $Y$-polarization since we are primarily interested in TE-modes. Depending on the layer over which the integration is conducted, it might be convenient to restore the $z$-dependence of the fields either from their values in the middle of the layer or on its boundary. Therefore, we calculate the required integrals for various limits that might be needed:
\begin{multline}
    \left.\int_{-h}^h[\mathbf{E}_1\mathbf{H}_2]\hat{x}dz\right|^y =
         \int_{-h}^h
         [(A_{y1}^+e^{ik_{z1}z}+A_{y1}^-e^{-ik_{z1}z})\frac{k_{x2}}{k_0}(A_{y2}^+e^{ik_{z2}z}+A_{y2}^-e^{-ik_{z2}z})]e^{i(k_{x1}+k_{x2})x}dz=\\
         =e^{i(k_{x1}+k_{x2})x}\frac{k_{x2}}{k_0}
         [\frac{1}{i(k_{z1}+k_{z2})}(e^{i(k_{z1}+k_{z2})h}-e^{-i(k_{z1}+k_{z2})h})A_{y1}^+A_{y2}^++\frac{1}{-i(k_{z1}+k_{z2})}(e^{-i(k_{z1}+k_{z2})h}-e^{i(k_{z1}+k_{z2})h})A_{y1}^-A_{y2}^-+\\+\frac{1}{i(k_{z1}-k_{z2})}(e^{i(k_{z1}-k_{z2})h}-e^{-i(k_{z1}-k_{z2})h})A_{y1}^+A_{y2}^-+\frac{1}{-i(k_{z1}-k_{z2})}(e^{-i(k_{z1}-k_{z2})h}-e^{i(k_{z1}-k_{z2})h})A_{y1}^-A_{y2}^+]=\\
         =e^{i(k_{x1}+k_{x2})x}\frac{k_{x2}}{k_0}
         \left[2\frac{\sin[(k_{z1}+k_{z2})h] }{k_{z1}+k_{z2}}\left(A_{y1}^+A_{y2}^++A_{y1}^-A_{y2}^-\right)+2\frac{\sin[(k_{z1}-k_{z2})h]}{k_{z1}-k_{z2}}\left(A_{y1}^+A_{y2}^-+A_{y1}^-A_{y2}^+\right)\right]
\end{multline}

\begin{multline}
    \left.\int_{-h}^0[\mathbf{E}_1\mathbf{H}_2]\hat{x}dz\right|^y =
         \int_{-h}^0
         [(A_{y1}^+e^{ik_{z1}z}+A_{y1}^-e^{-ik_{z1}z})\frac{k_{x2}}{k_0}(A_{y2}^+e^{ik_{z2}z}+A_{y2}^-e^{-ik_{z2}z})]e^{i(k_{x1}+k_{x2})x}dz=\\
         =e^{i(k_{x1}+k_{x2})x}\frac{k_{x2}}{k_0}
         [\frac{1}{i(k_{z1}+k_{z2})}(1-e^{-i(k_{z1}+k_{z2})h})A_{y1}^+A_{y2}^++\frac{1}{-i(k_{z1}+k_{z2})}(1-e^{i(k_{z1}+k_{z2})h})A_{y1}^-A_{y2}^-+\\+\frac{1}{i(k_{z1}-k_{z2})}(1-e^{-i(k_{z1}-k_{z2})h})A_{y1}^+A_{y2}^-+\frac{1}{-i(k_{z1}-k_{z2})}(1-e^{i(k_{z1}-k_{z2})h})A_{y1}^-A_{y2}^+]=\\
         =e^{i(k_{x1}+k_{x2})x}\frac{k_{x2}}{k_0}
         \left[2\frac{\sin[(k_{z1}+k_{z2})h/2] }{k_{z1}+k_{z2}}\left(A_{y1}^+A_{y2}^+e^{-i({k_{z1}+k_{z2})h/2}}+A_{y1}^-A_{y2}^-e^{i({k_{z1}+k_{z2})h/2}}\right)+\right.\\\left.+2\frac{\sin[(k_{z1}-k_{z2})h/2]}{k_{z1}-k_{z2}}\left(A_{y1}^+A_{y2}^-e^{-i({k_{z1}-k_{z2})h/2}}+A_{y1}^-A_{y2}^+e^{i({k_{z1}-k_{z2})h/2}}\right)\right]
\end{multline}

For the boundary layers we obtain simpler expressions:

\begin{multline}
    \int_{0}^\infty[\mathbf{E}_1\mathbf{H}_2]\hat{x}dz
         =e^{i(k_{x1}+k_{x2})x}\frac{k_{x2}}{k_0}
         \frac{1}{i(k_{z1}+k_{z2})}(0-1)A_{y1}^+A_{y2}^+
         =-e^{i(k_{x1}+k_{x2})x}\frac{k_{x2}}{k_0}
         \frac{1}{i(k_{z1}+k_{z2})}A_{y1}^+A_{y2}^+
\end{multline}

\begin{equation}
    \int_{-\infty}^0[\mathbf{E}_1\mathbf{H}_2]\hat{x}dz
         =-e^{i(k_{x1}+k_{x2})x}\frac{k_{x2}}{k_0}
         \frac{1}{i(k_{z1}+k_{z2})}A_{y1}^-A_{y2}^-
\end{equation}

Knowing these expressions for integrals, we can easily calculate the amplitude of the mode according to Eq.~\ref{Sup:amplitude}. Finally, in order to calculate the flux of the excited $s$-mode, $S_s^{\mathrm{wg}}$, we need to calculate the Poynting vector flux for the "unitary" mode, $S_s^{\mathrm{wg,0}}$:

\begin{equation}
    S_s^{\mathrm{wg}}=|C_s|^2 S_s^{\mathrm{wg},0}.
\end{equation}

In this way the Poynting flux $S_s^{\mathrm{wg},0} = \frac{c}{8\pi}\int[\mathbf{E}_s\mathbf{H}^*_s]d\mathbf{S}$ might be calculated via the following integrals derived in the terms introduced above:
\begin{multline}
    \frac{c}{8\pi}\int[\mathbf{E}\mathbf{H}^*]\hat{x}dz = \frac{c}{8\pi}\int
         (E_yH^*_z-E_zH^*_y)dz=
         \frac{c}{8\pi}\int
         (A_{y}^+e^{ik_{z}z}+A_{y}^-e^{-ik_{z}z})\frac{k_{x}}{k_0}(A_{y}^{+*}e^{-ik^*_{z}z}+A_{y}^{-*}e^{ik^{*}_{z}z})dz=\\=\frac{c}{8\pi}\frac{k_{x}}{k_0}\int
         |A_{y}^+|^2e^{-2\mathrm{Im} k_{z}z}+|A_{y}^-|^2e^{2\mathrm{Im} k_{z}z}+A_y^+A_y^{-*}e^{2i\mathrm{Re}k_z z}+A_y^-A_y^{+*}e^{-2i\mathrm{Re}k_z z}dz
\end{multline}

\begin{multline}
    \frac{c}{8\pi}\int_{-h}^h[\mathbf{E}\mathbf{H}^*]\hat{x}dz = \frac{c}{8\pi}\int_{-h}^h
         (E_yH^*_z-E_zH^*_y)dz=\\=\frac{c}{8\pi}\frac{k_{x}}{k_0}\left[
         \frac{-2\mathrm{sh}(2\mathrm{Im}k_zh)}{-2\mathrm{Im}k_z}|A_{y}^+|^2+\frac{2\mathrm{sh}(2\mathrm{Im}k_zh)}{2\mathrm{Im}k_z}|A_{y}^-|^2+\frac{2i\mathrm{sin}(2\mathrm{Re}k_zh)}{2i\mathrm{Re}k_z}A_y^+A_y^{-*}+\frac{-2i\mathrm{sin}(2\mathrm{Re}k_zh)}{-2i\mathrm{Re}k_z}A_y^-A_y^{+*}\right]=\\
         =\frac{c}{8\pi}\frac{k_{x}}{k_0}\left[
         \frac{\mathrm{sh}(2\mathrm{Im}k_zh)}{\mathrm{Im}k_z}\left(|A_{y}^+|^2+|A_{y}^-|^2\right)+\frac{\mathrm{sin}(2\mathrm{Re}k_zh)}{\mathrm{Re}k_z}\left(A_y^+A_y^{-*}+A_y^-A_y^{+*}\right)\right]=
\end{multline}

\begin{multline}
    \frac{c}{8\pi}\int_0^\infty[\mathbf{E}\mathbf{H}^*]\hat{x}dz = \frac{c}{8\pi}\int_0^\infty
         (E_yH^*_z-E_zH^*_y)dz=\frac{c}{8\pi}\frac{k_{x}}{k_0}\frac{1}{-2\mathrm{Im}k_z}(0-1)
         |A_{y}^+|^2=\frac{c}{8\pi}\frac{k_{x}}{k_0}\frac{1}{2\mathrm{Im}k_z}
         |A_{y}^+|^2
\end{multline}

\begin{multline}
    \frac{c}{8\pi}\int_{-\infty}^0[\mathbf{E}\mathbf{H}^*]\hat{x}dz = \frac{c}{8\pi}\int_{-\infty}^0
         (E_yH^*_z-E_zH^*_y)dz=\frac{c}{8\pi}\frac{k_{x}}{k_0}
         \frac{1}{2\mathrm{Im}k_z}|A_{y}^-|^2(1-0)=\frac{c}{8\pi}\frac{k_{x}}{k_0}
         \frac{1}{2\mathrm{Im}k_z}|A_{y}^-|^2
\end{multline}

\end{widetext}

%\bibliography{sample}

%apsrev4-2.bst 2019-01-14 (MD) hand-edited version of apsrev4-1.bst
%Control: key (0)
%Control: author (8) initials jnrlst
%Control: editor formatted (1) identically to author
%Control: production of article title (0) allowed
%Control: page (0) single
%Control: year (1) truncated
%Control: production of eprint (0) enabled
%

\end{document}